\documentclass{IEEEcsmag}

\usepackage[colorlinks,urlcolor=blue,linkcolor=blue,citecolor=blue]{hyperref}

\usepackage{upmath}
\usepackage{tabularx}
\usepackage[table]{xcolor}
\usepackage{graphicx}

\newcommand{\toolName}{\textit{SpaudioData}}

\newcommand{\revision}[1]{\leavevmode{\textcolor[RGB]{220, 20, 60}{#1}}}
\newcommand{\remark}[1]{\textcolor[RGB]{150, 200, 0}{#1}}

\newcommand{\revisionminor}[1]{\leavevmode{\textcolor[RGB]{220, 20, 60}{#1}}}

\newcommand{\removed}[1]{\leavevmode{\color{red}{\st{#1}}}}

\def \cleanversion{} 
\ifx\cleanversion\undefined
\else
 \renewcommand{\revisionminor}[1]{#1}
 \renewcommand{\remark}[1]{\iffalse #1 \fi} 
 \renewcommand{\removed}[1]{\iffalse #1 \fi} 
 \renewcommand{\revision}[1]{#1}
\fi

\begin{document}


\title{Accessible Fine-grained Data Representation
via Spatial Audio}

\author{
Can Liu\textsuperscript{\dag},
Wenjie Jiang\textsuperscript{\dag}
}
\affil{Nanyang Technological University, Singapore
}

\author{
Shaolun Ruan,
Kotaro Hara}
\affil{Singapore Management University, Singapore}

\author{
Yong Wang\textsuperscript{\S}}
\affil{Nanyang Technological University, Singapore}


\begin{abstract}
\looseness-1
Pitch-based sonification of quantitative data increases the accessibility of data visualizations that are otherwise inaccessible for blind and low-vision (BLV) individuals.
We argue that, although pitch representations can reveal the coarse-grained information of data, such as data trend and value comparison, they cannot effectively convey the fine-grained details like the sign and exact value of individual data points.
Informed by existing sound perception research, we propose a spatial audio-based approach by representing data values as the sound direction in the \revision{azimuth} plane to achieve accessible fine-grained data representation.
We conducted a user study with \revision{26} participants (including \revision{10} BLV participants) on four data perception tasks. 
The results show our approach significantly outperforms pitch representation on fine-grained data perception tasks like recognizing data signs and exact values, and performs similarly on data trend identification, despite its inferior accuracy on data value comparison.


\end{abstract}

\maketitle

\renewcommand{\thefootnote}{}
\footnotetext{\dag~C. Liu and W. Jiang contributed equally to this work.}
\footnotetext{\S~Y. Wang is the corresponding author.}

\renewcommand{\thefootnote}{\arabic{footnote}}

\section{Introduction}

Data visualization has been commonly used in various applications with the growing data-driven nature of our society. 
However, data visualizations leverage human visual perception to convey the underlying data, which makes them intrinsically inaccessible to people with visual impairment. 
To solve this issue, the data visualization and accessibility fields have developed a series of approaches to enhance the accessibility of data visualizations for blind and low-vision (BLV) people in the past few years~\cite{kim2021accessible,siu2022supporting}.
Sonification, the process of converting data into sound, has emerged as a promising alternative method for data representation to increase data visualization accessibility. 
For example, by converting input data into audio information (\textit{e.g.,} pitch), sonification allows BLV users to discern the overall trends of the corresponding data auditorily~\cite{siu2022supporting}.
Despite the availability of different
audio
features (\textit{e.g.,} pitch, timbre and loudness), existing approaches typically employ pitch (\textit{i.e.,} the frequency of sound) to represent data values~\cite{kim2021accessible}.

According to our survey of existing pitch-based approaches, pitch has been shown to be effective in conveying \textbf{coarse-grained data information}, such as data trend analysis and value comparison~\cite{siu2022supporting}, as human users are adept at perceiving differences in pitch.
But pitch seems ineffective in conveying the \textbf{fine-grained data details} (\textit{e.g.,} the sign and exact value of individual data points), as it is challenging for users without a background of professional acoustics or musical training to identify the \emph{exact frequency value} (in \revision{Hz}) of a sound.
It is also the reason why pitch is often combined with speech-based and touch-based approaches, where detailed data values are conveyed via speech or touch on demand~\cite{siu2022supporting}. 
These combinations by nature require BLV users to conduct extra interactions and make the data exploration less efficient.
Motivated by this, the paper aims to answer one critical research question: \textbf{\textit{Can we achieve accessible data representation with perceptible fine-grained details via sonification without speech-based verbalization?}}


In this paper, we develop a novel sonification technique that uses spatial audio to encode the underlying data, aiming to achieve accessible and fine-grained data representation. The development of our approach is inspired by the existing applications and extensive research on spatial audio, as well as the easy accessibility of spatial audio devices. First, spatial audio, which is often realized by using the head-related transfer functions (HRTFs), has been widely used in virtual reality or augmented reality systems~\cite{yang2022audio}.
Second, there have been extensive studies on spatial audio in the perception field~\cite{rebillat2012audio,steadman2019short}. These studies provide the fundamental perception guidelines for the design of our approach.
Third, spatial audio is commonly supported by many commercial earphones and headphones, such as Apple AirPods\footnote{\url{https://www.apple.com/sg/airpods-4/}}, Sony WH-1000XM5 headphone\footnote{\url{https://www.sony.com.sg/electronics/headband-headphones/wh-1000xm5}} and Samsung Galaxy Buds Pro\footnote{\url{https://www.samsung.com/sg/audio-sound/galaxy-buds/galaxy-buds-pro-black-sm-r190nzkaxsp/}}.
\revisionminor{This widespread hardware support suggests that spatial audio–based approaches could be deployed on commonly available consumer devices.}



Specifically, informed by prior perception research on localization accuracy of sound direction~\cite{steadman2019short}, we propose leveraging sound direction in the \revision{azimuth} plane (horizontal plane) of the virtual 3D environment to indicate the data values, as illustrated in Figure \ref{fig:method_illus}.
\revisionminor{In this design, the frontal direction corresponds to zero, enabling negative and positive values to be represented by spatial deviations to the left and right.}
Given the multiple variables in the design space of our approach, 
we first conducted a pilot study to determine important design variables, including angle interval and sound repetition. 
We then performed a user study to compare our approach with pitch representation, the most widely-used sonification approach, on
both coarse-grained data exploration tasks involving multiple numbers (\textit{i.e.,} sequential data trend identification and two-value comparison) and fine-grained tasks involving a single number (\textit{i.e.,} sign identification and exact value recognition).
We recruited \revision{26} participants (\revision{10} BLV participants) in our user study.
The user study results demonstrate our approach's effectiveness in achieving accessible fine-grained data representation for both sighted and BLV people. 
In particular, we have three major findings: 1) our spatial audio approach significantly outperforms pitch representation in the data trend identification task and the fine-grained tasks of sign identification and exact value recognition; and 2) the pitch representation is more accurate for participants to perform numeric value comparisons than our spatial audio approach.
We discuss the detailed implications of these findings and the lessons learned from developing our approach.

In summary, our contributions are as follows:
\begin{itemize}
    \item We show that spatial audio can be harnessed to explicitly represent numeric data with fine-grained details;
    \item We propose \toolName{}, a \underline{sp}atial \underline{audio}-based approach for accessible \underline{data} representation. It employs the sound direction in the \revision{azimuth} plane to encode numeric data, which is the first of its kind work. The implemented desktop application and the source code of our approach have been released at
    \textcolor{blue}{\textit{[https://github.com/orgs/SpaudioData/repositories].}}
    
    
    \item We conduct an extensive user study with \revision{26} participants (\revision{10} BLV participants) to evaluate the effectiveness and usability of our approach. The results confirm that \toolName{} significantly outperforms the widely-used pitch representation in fine-grained tasks, and has a comparable or slightly lower accuracy than pitch representation on coarse-grained tasks.

\end{itemize}

\section{Related Work}

Our work is related to prior research on
accessible data representation, data sonification, and spatial audio application and perception.

\subsection{Accessible Data Representation}

Previous research~\cite{kim2021accessible} on accessible data representation has explored utilizing auditory, tactile, and even olfactory senses to convey information to BLV individuals. 
However, many such approaches, particularly those involving olfactory or specialized haptic feedback, often require dedicated hardware that limits their applicability in everyday scenarios.
A more flexible and scalable solution involves using alternate texts in browsers; however, BLV users often encounter significant challenges in obtaining accurate information efficiently through static text alone~\cite{sharif2021understanding}.

To move beyond passive or static data consumption, recent research has focused on interactive and authorable non-visual representations. 
Sharif et al.~\cite{sharif2022voxlens} developed VoxLens, a JavaScript plugin that provides an interactive way for BLV individuals to access online data, significantly improving accuracy and reducing interaction time. 
\revisionminor{Similarly, to empower BLV users to become creators of their own data experiences, Potluri et al.~\cite{potluri2022psst} introduced PSST, a toolkit that enables BLV developers to author and customize sonifications for real-time streaming sensor data. }
Building on these interactive non-visual approaches, data sonification provides a robust theoretical and practical framework for representing complex datasets.

\subsection{Data Sonification}

\revisionminor{Drawing on established visualization theory, sonification techniques map data onto a temporal substrate through the use of auditory marks, encoding information into attributes like pitch, timbre, and loudness as auditory channels to form coherent auditory graphs~\cite{enge2024open}.}
These techniques allow for the exploration of data in a non-visual format, enabling users to perceive patterns, trends, and relationships through sound.
Building on this foundational understanding, numerous studies have explored specific encoding methods that offer distinct advantages for different data types and tasks.
Statistical analysis of existing research~\cite{muchnik1991central} shows that pitch is the most commonly utilized auditory attribute for encoding data, primarily due to its high resolution. However, this method is predominantly used in scenarios involving data trend analysis and comparison, where changes in pitch effectively highlight variations across data sets.
\revision{
Also, Holloway et al.~\cite{holloway2022infosonics} studied how to better support the BLV participants' perception of infographics using sound, i.e., sonification and voice.}
\revisionminor{Their work primarily focused on conveying infographic content through these auditory channels, while the use of spatial audio direction to represent numerical values was not explored.}

\revisionminor{Among the sonification approaches, a few have incorporated spatial audio to facilitate accessible data exploration~\cite{nasir2007sonification}.}
However, spatial audio has been primarily used in an auxiliary capacity, such as for navigation purposes. These studies have not yet explored the potential of using spatial audio to encode numeric data explicitly.
This gap in research underscores the necessity for our work, which aims to extend the capabilities of spatial audio beyond mere navigational aids and into a robust tool for encoding and understanding complex numerical data sets. This approach could significantly enhance data accessibility and interpretation, particularly for the BLV community.
\revision{
In addition to this Wang et al.~\cite{wang2022seeing} investigated how different auditory channels map to data and chart types, 
However, our paper goes beyond comparative guidelines by proposing and validating spatial audio direction as a novel encoding channel that explicitly supports fine-grained numeric perception.}

\subsection{Spatial Audio Application and Perception}

In the field of spatial audio for assisting individuals with visual impairments, prior research has primarily focused on three areas: navigation and wayfinding, exploration of virtual environments, and interaction with maps and geographical information.
Navigation and wayfinding studies have focused on the development of systems to guide visually impaired users through real-world environments. Kammoun et al.~\cite{kammoun2012navigation} described the NAVIG project, which aims to assist navigation and space perception for visually impaired users through a combination of spatial audio and other sensory modalities.
Prior research on the exploration of virtual environments has led to the development of novel spatial sonification techniques for non-visual interaction.
Geronazzo et al.~\cite{geronazzo2016interactive} explored interactive spatial sonification for non-visual exploration of virtual maps, demonstrating how spatial audio can enhance the understanding of spatial layouts.
The third category of studies focuses on the interaction with maps and geographical information.


The above research and application of spatial audio have shown the power of spatial audio, which motivates us to explore the usage of spatial audio for accessible fine-grained data representation.

\section{Background}
Before introducing the detailed design of our spatial audio-based approach, we \revision{present} the relevant background knowledge of our approach in this section.

\textbf{Spatial Audio.}
Spatial audio refers to the simulation of sound in three-dimensional space, enabling listeners to perceive sound sources as being located at specific positions in space relative to their own position. The key feature of spatial audio is its ability to convey directional cues, such as where a sound is coming from, its distance from the listener, and even its movement.
A common approach to modeling spatial audio is through the use of spherical coordinates ~\cite{letowski2011localization}, with the listener's head positioned at the center.
Based on spatial audio, one of the key factors in data representation is the listener's sound localization ability. The azimuth plane and elevation plane are crucial concepts in this context. The azimuth plane refers to the horizontal plane, where sound sources are localized in terms of their angular position around the listener, typically ranging from 0° to 360°. On the other hand, the elevation plane represents the vertical dimension, where sound sources are localized above or below the listener's head, with angles typically ranging from -90° (directly below) to +90° (directly above).
The process of sound localization can be seen as a combination of determining the direction of sound in both the azimuth and elevation planes. This two-dimensional framework allows the listener to perceive and identify the exact spatial location of a sound source in three-dimensional space.
\revision{
Specifically for our approach, we leverage this framework to map numerical data values to the azimuth plane, taking advantage of the human auditory system's higher localization accuracy in the horizontal direction.
}

\textbf{Frequency vs. Pitch.}
Frequency is the number of sound vibrations or cycles per second, measured in Hertz (Hz). It is a physical property of the sound wave and is the primary determinant of the pitch of the sound.
Pitch is the human perception of the frequency of sound. It is how high or low a sound appears to a listener. Pitch is not a direct property of the sound itself but a subjective sensation that depends on the frequency as well as the listener’s auditory system. For instance, a higher frequency generally corresponds to a higher perceived pitch.

\textbf{Head-Related Transfer Functions (HRTFs).}
The Head-Related Transfer Functions (HRTFs) are a crucial concept in spatial audio processing, used to describe how sound is transmitted to the ears through the human head, ears, and body, and how these anatomical structures influence the sound's perception. In simple terms, HRTFs model the relationship between the position of a sound source in the external environment and the sound signals that reach the eardrums, influenced by the listener's unique physical features.
HRTFs are fundamental for sound localization, helping listeners perceive the direction of sounds. By analyzing differences in arrival times (ITD, Interaural Time Difference), intensity (ILD, Interaural Level Difference), and frequency characteristics (such as filtering effects from the pinna), the human brain is able to determine the position of the sound source.
An important feature of HRTFs is the individual variation between people. Each person’s head, ears, and body are anatomically distinct, making their HRTFs unique as well. Steadman et al.~\cite{steadman2019short} pointed out that personalized HRTFs models can be used to provide the most accurate spatial audio experience, but due to these individual differences, most current audio systems still rely on generalized HRTFs models. This can lead to inaccuracies in sound localization, particularly in more complex spatial dimensions like height and distance.
\revision{We explicitly considered this challenge in the design and implementation of \toolName{} by leveraging the generic HRTF models provided by commercial hardware (i.e., Apple AirPods).}

\section{Spatial Audio-Based Data Representation}
\revision{
Building upon the principles of spatial audio and HRTFs described in the previous section, we explore the possibility of leveraging spatial audio to achieve accessible fine-grained data representation for BLV people. Our design specifically addresses the perceptual characteristics discussed in the background.
}

\subsection{Design Considerations}
In designing sonification systems, it is essential to account for various factors that influence how auditory information is perceived by users. These include the challenges of accurately judging egocentric distance and the inherent differences in sound localization across spatial planes. This section discusses two critical considerations: (1) the perceptual biases associated with egocentric distance and (2) the differences in localization accuracy between the \revision{azimuth} and elevation planes. By carefully addressing these issues, we propose using sound directions in the \revision{azimuth} plane to effectively encode numeric data.

\textbf{Egocentric distance vs. sound direction}. 
Due to the phenomenon of distance compression~\cite{rebillat2012audio}, participants often struggle with accurately judging distances. Specifically, they tend to underestimate close distances while overestimating far ones. As a result, using distance to convey precise numerical data may lead to confusion or misinterpretation.
To address this issue, we opt to use sound direction rather than distance to encode data. Sound direction is perceived more consistently and with greater accuracy, as humans are generally better at determining the direction of a sound source than its distance. 

\textbf{Azimuth plane vs. elevation plane}.
Prior research~\cite{letowski2011localization} has demonstrated that humans are significantly more accurate at localizing sound in the \revision{azimuth} plane (the horizontal plane) than in the elevation plane (the vertical plane). This difference in localization accuracy stems from the way the auditory system processes spatial information. In the \revision{azimuth} plane, the brain can rely on interaural time differences and interaural level differences, both of which provide highly precise cues for sound localization. These cues are especially effective in the horizontal plane because sound waves from a source in this direction reach each ear at slightly different times and levels, which allows the brain to accurately infer the direction of the sound source.
In contrast, the elevation plane poses greater challenges for auditory localization. The primary cues for vertical localization come from the spectral shape of the sound, which is influenced by the shape of the outer ear (pinna). However, these cues are much less reliable and vary depending on the frequency content of the sound and the listener's head position. As a result, our ability to localize sounds in the vertical plane is generally less precise. Given these findings, we propose focusing on the \revision{azimuth} plane for our approach.
By collectively considering the factors from the above two perspectives, we propose using sound directions in the \revision{azimuth} plane to encode the number.

\subsection{Sound Direction Encoding}
As demonstrated by the prior study~\cite{steadman2019short}, the sound localization error at the back of human users is relatively high, therefore, we can use the directions of the front semicircle to represent the data.

\begin{figure}
    \centerline{\includegraphics[width=\columnwidth]{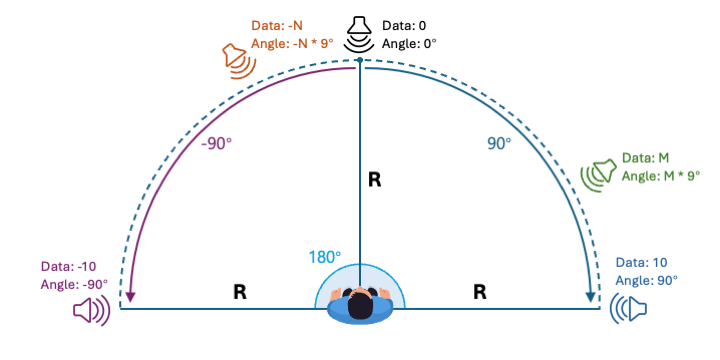}}
    \caption{Illustration of data encoding in \toolName. The encoding space forms a semicircle in front of the participant, spanning \(180^\circ\). Sounds from the left, straight ahead, and right represent data -10, 0, and 10 respectively. Negative values are encoded as leftward spatial audio, while positive values are encoded as rightward. The angular interval is \(9^\circ\).}
    \label{fig:method_illus}
\end{figure}

For each data point, we project the value onto the arc with the radius $R = 3$ meters in virtual space. A mapping function was used to calculate from the low range to the high range of the data (in our case, we used -10 to 10
and linearly mapped the value from this range to a $\theta$ in the range of \(-90^\circ\) to \(90^\circ\). This range was selected to balance between perceptual discriminability in azimuth space and comparability to pitch-based method, while avoiding perceptual overload in fine-grained directional tasks.).

The virtual coordinate location of the data point will then be projected using a polar-to-cartesian coordinate translation function to calculate the destination location. In order to clarify the method we used for projection, we have listed the function below.

\begin{equation}
\label{equation:1}
\begin{split}
& x = R \cdot \cos{\theta},\\
& y = R \cdot \sin{\theta},\\
\end{split}
\end{equation}

\noindent where $\theta$ is the angle of the sound source in the polar system, and $x$ and $y$ are the corresponding coordinates of the sound source in the 2D \revision{azimuth} plane. 
Figure~\ref{fig:method_illus} illustrates the details of our encoding design.

\subsection{Implementation}
\revision{To validate our design concepts and evaluate the proposed mapping strategy,}
we implemented our approach as a web-based application, \toolName{}. It consists of the following modules: data configuration user interface, spatial audio generation, and headphone choice.

\textbf{Data Configuration User Interface}.
To ensure that the system is accessible to individuals who are visually impaired, a user interface was designed and optimized for sound-based interaction. Accessible components were added to serve such purposes, such as voice-over of the value during the training progress. For debugging purposes, a simple visualization tool was created to display the correlation of spatial audio and the mapping of the data. 


\textbf{Spatial audio generation}.
We employed Faust Live\footnote{\url{https://faust.grame.fr}} as the core software to generate spatial audio. Specifically, we implemented spatial audio effects using the Faust programming language. This system utilizes spherical harmonics to encode and decode audio signals, enabling precise simulation of sound locations within a three-dimensional space.

\textbf{\revision{Headphone} choice}.
\revision{McMullen}~\cite{mcmullen2014potentials} discussed that, compared to external speakers, headphones provide a better spatial audio experience because they can maintain a fixed position relative to the listener. Another point is that, although sound localization can be achieved through various means, such as 
Dolby Atmos for the Home\footnote{https://professional.dolby.com/tv/home/dolby-atmos/},
to ensure accessibility and portability, we will use headphones as the medium for delivering binaural audio. In our experiment, we selected Apple AirPod as the headphone, 
\revision{because 1) it enables an immersive spatial audio experience, and 2) more importantly, it provides generic HRTF models without the need of complex personalized calibration, which can minimize the perceptual errors associated with non-personalized HRTFs and make \toolName{} accessible to a broad range of users.
}

\section{Pilot Study}
\revision{Before conducting the formal comparative evaluation, it was necessary to determine the optimal configuration for our spatial encoding parameters.}
In the preliminary phase of our research, we tested several parameters of the \toolName{} on 6 sighted participants. These parameters included the interval angle and the number of sound repetitions, which are crucial for optimizing system performance.

\subsection{Participant}

We recruited 6 university students (three males and three females) aged between 23 and 26 years. All participants possessed basic mathematical skills, understanding positive and negative numbers and magnitude comparisons, and had no visual impairments. One participant had a background in music training, but none had prior experience with sonification technologies. Each participant received compensation of 15 Singapore dollars upon completing the experiment.
The study protocol was reviewed and approved by our university’s Institutional Review Board.

\subsection{Study Procedure}

\revision{
All experiments were conducted in a quiet indoor setting with only one participant and one study coordinator present at a time. Upon arrival, participants signed an informed consent form prior to the experiment.
The study coordinator assisted them in wearing headphones and adjusting the volume to a comfortable level.
Participants were then instructed to familiarize themselves with the experimental goal.
The primary task was a single-value guessing task, in which participants listened to a sonified representation of a number and were asked to verbally state the estimated value to the study coordinator.
The total duration of the pilot study was approximately 30 minutes per participant.
Each trial had a one-minute time limit to ensure consistent pacing.
Following the testing phase, a brief interview was conducted to gather qualitative feedback.
After the experiment, all participants completed a demographic survey covering their age, gender, vision level, education, musical background, and hearing status.}
The experiment was designed with two interval angles (3° and 9°) and three sound repetition frequencies (1, 3, and 5 times), resulting in 6 test combinations. A single-value guessing task was selected, randomly choosing 10 numbers from the range of -10 to 10.


\subsection{Findings}

\begin{figure}
    \centerline{\includegraphics[width=\columnwidth]{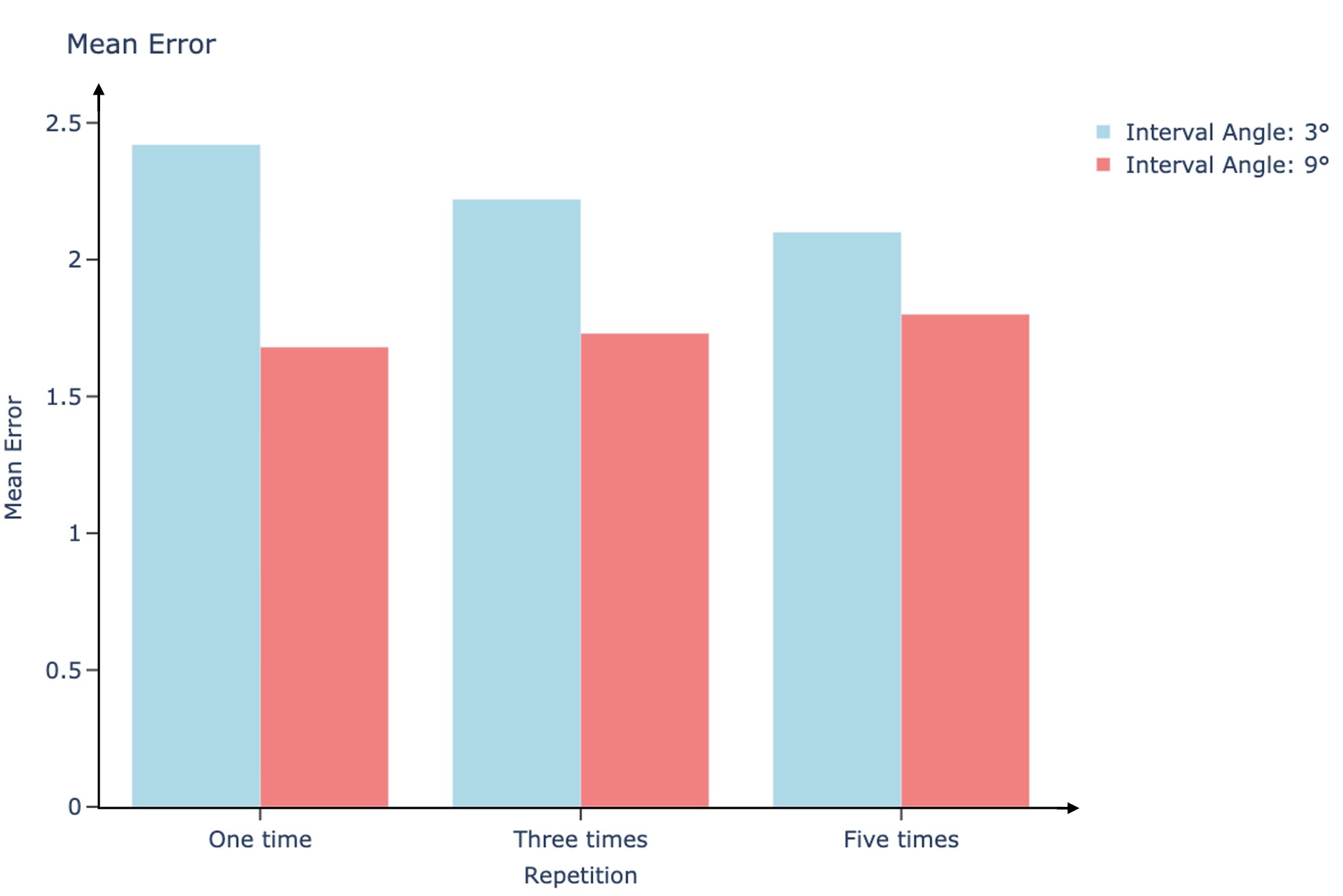}}
    \caption{\textbf{Pilot Study Results.} The mean error for the \(9^\circ\) interval angle was smaller than that observed for the \(3^\circ\) configuration. For the \(9^\circ\) condition, errors for the one-playback and three-playback conditions were similar and lower than those in the five-playback condition.}
    \label{fig:pilot_study_result}
\end{figure}

\autoref{fig:pilot_study_result} shows the experimental results, with the average error computed for each combination.

\subsubsection{Interval Angle.}
All participants reported that an interval angle of \(9^\circ\) allowed them to distinguish between sounds more clearly. Experimental data also indicated that the average error was significantly lower with the \(9^\circ\) interval angle compared to \(3^\circ\).

\subsubsection{Sound Repetition.}
Most participants noted that they often missed sounds when they were played only once, while five playbacks led to auditory fatigue.
Experimental results demonstrated that a setting of three playbacks minimized errors most effectively.
These results were used to determine the default playback configuration of \toolName{} in the formal study.

\section{Formal Study}
To further assess the utility and effectiveness of \toolName{}, we conducted a user study that compared spatial audio with methods that use pitch changes to convey information.
\revision{The user study was conducted in accordance with the ethical standards of the university. Ethical approval was obtained from the Institutional Review Board of our university before the commencement of the study. All participants provided informed consent prior to participation.}

\subsection{Participants}
\revision{10 BLV participants and 16 sighted participants} participated in this study, none of whom had significant prior experience with sonification.
Out of the 26 participants, two reported having a professional music training background. None reported any hearing loss.

\begin{table}
\setlength{\tabcolsep}{4pt}
\caption{Details of Participants (``Exp.'' refers to the participant's musical training experience).}
\label{tab:partici_details}
\centering
\rowcolors{2}{gray!25}{white}
\begin{tabularx}{\columnwidth}{cccccX}
\hline
\rowcolor{gray!50}
\textsc{ID} & \textsc{Gender} & \textsc{Age} & \textsc{Vision Level} & \textsc{Education} & \textsc{Exp.} \\
\hline
P1 & Male & 26 & Sighted & Master & None \\
P2 & Female & 23 & Sighted & Bachelor & None \\
P3 & Male & 25 & Sighted & Bachelor & None \\
P4 & Male & 28 & Sighted & Master & Train \\
P5 & Female & 28 & Sighted & Master & None \\
P6 & Female & 24 & Sighted & Bachelor & Train \\
P7 & Female & 24 & Sighted & Bachelor & None \\
P8 & Female & 27 & Sighted & Master & None \\
P9 & Male & 24 & Sighted & Bachelor & None \\
P10 & Male & 27 & Sighted & Bachelor & None \\
P11 & Female & 26 & Sighted & Bachelor & None \\
P12 & Female & 24 & Sighted & Bachelor & None \\
P13 & Female & 27 & Sighted & Master & None \\
P14 & Female & 26 & Sighted & Master & None \\
P15 & Male & 26 & Sighted & Bachelor & None \\
P16 & Male & 24 & Sighted & Bachelor & None \\
P17 & Male & 53 & Blind & High School & None \\
P18 & Female & 60 & Blind & Bachelor & None \\
P19 & Male & 40 & Low Vision & Bachelor & None \\
P20 & Female & 55 & Blind & High School & None \\
P21 & Male & 60 & Blind & High School & None \\
P22 & Female & 56 & Blind & Bachelor & None \\
P23 & Male & 28 & Low Vision & Bachelor & None \\
P24 & Female & 51 & Low Vision & High School & None \\
P25 & Male & 56 & Low Vision & High School & None \\
P26 & Female & 48 & Blind & Bachelor & None \\
\hline
\end{tabularx}
\end{table}

\subsection{Task Design}

\revision{We conducted a controlled laboratory study using a within-subjects design.
Each participant completed all four tasks under two experimental conditions: 
(1) using \toolName{} and (2) using the baseline pitch method.
To extensively evaluate the performance of \toolName{}, the four tasks are designed to cover both \textit{coarse-grained data information} (comparison and trend) and \textit{fine-grained data details}.}
The coarse-grained data exploration tasks provide users with a basic gist overview of the data, and have been commonly covered in prior studies on using sonification for accessible data representation~\cite{siu2022supporting}. Specifically, we mainly consider two coarse-grained data exploration tasks:

\begin{enumerate}
    \item \textbf{Data value comparison:}
    Data value comparison is a fundamental task in data exploration. For example, in bar charts and line charts, users need to conduct value comparison to identify the outliers such as the maximum and minimum numbers. This task has also been investigated by prior data accessibility research~\cite{hoque2023accessible}. In our formal study, participants are asked to compare 
    two integer numbers
    ranging from -10 to 10 and determine which one is greater.
    
    \item \textbf{Data trend recognition:} When conducting data exploration, it is crucial to recognize the overall trend of the underlying data to gain a \revision{general} overview.
    For example, a primary purpose of line charts is to show the overall trend of time-series data. This type of task has also been widely investigated in prior accessibility studies~\cite{siu2022supporting}.
    In our study, participants are asked to identify the trend of five integer numbers, including four common types: monotonically increasing, monotonically decreasing, increasing then decreasing, or decreasing then increasing.
\end{enumerate}

While coarse-grained tasks provide a broad overview, fine-grained tasks are necessary for precise data interpretation.

\begin{enumerate}
    \item \textbf{Value sign identification:} 
    Identifying the sign of a number is a fundamental task in data exploration. Recognizing whether a value is positive or negative is crucial for tasks that involve distinguishing between two categories. In this study, participants are asked to determine if a given number, ranging from -10 to 10, is positive, negative, or zero.
    \item \textbf{Exact value identification:} 
    The previous tasks focus on coarse-grained distinctions, whereas exact value identification requires finer-grained perception. Participants were asked to precisely identify the numerical value of a given number.
\end{enumerate}

By including both coarse- and fine-grained tasks, we can compare \toolName{} with prior studies and highlight its unique encoding capabilities, providing insights into how data granularity affects performance.

\subsection{Dataset}
\revision{We carefully designed the numbers conveyed via audio for each task. All values were integers within the range $[-10, 10]$.}

\revision{For the data value comparison task, we created 12 pairs of numbers, with their differences being $0, 3, 6,$ and $9$ to systematically vary the difficulty of tasks. In general, a smaller value difference (e.g., $0$ and $3$) required finer auditory discrimination and makes it more difficult for participants to compare them correctly via audio, and a larger value difference (e.g., $9$) makes it relatively easier.
Such a systematic design of the testing data allows for a detailed analysis of how differences in numerical values may influence participants' performance in data value comparison via the audio.
The order of the whole set of number pairs was randomized in our study, avoiding possible bias resulted from the ordering effect.
}


\revision{For the task of data trend recognition, we asked participants to identify the trend of five numbers. We designed 12 sets of five numbers with an equal interval between two adjacent numbers. The interval ranges cover \([3, 6, 9]\) to progressively simplify the task, as a larger interval makes it easier for participants to identify the data trend via audio.
We considered}
four primary trends: monotonically increasing, monotonically decreasing, increasing then decreasing, and decreasing then increasing. Due to the range limits inherent in the intervals of 6 and 9, monotonically increasing and monotonically decreasing trends were intentionally excluded from these larger intervals to avoid exceeding the numerical bounds of -10 to 10. This ensures that all trends are represented within feasible numerical ranges, allowing for a balanced assessment of participants' abilities to discern and interpret these patterns. By randomizing the dataset's order, we prevent possible order effects that can influence the results.

\revision{
For the value sign identification and exact value identification tasks, we used the same set of data.
To ensure balanced sampling across the entire range, data were uniformly selected from the 21 integers between $-10$ and $+10$.
Consequently, the dataset consisted of 21 numbers, covering the full numerical spectrum.
During the study, participants were asked to first identify the sign of each value (positive, negative, or zero) and then report its exact value.}

\subsection{Baseline Method}
According to a recent study~\cite{hoque2023accessible}, pitch is the most commonly used encoding method in sonification.
The wide usage of pitch in sonification comes from its ability to span a wide range of frequencies, allowing it to convey both precise variations and broader trends.
Other auditory parameters, such as loudness, timbre, and tempo, can be useful in certain contexts, but they are generally less precise for encoding data~\cite{hu2020comparative}. Loudness lacks the resolution of pitch and is easily affected by environmental noise, making it generally unreliable for conveying exact values. Timbre is effective for distinguishing categories, but it does not provide the fine-grained variation needed for accurate data representation. Tempo, on the other hand, is typically used to convey urgency rather than specific data values. These limitations make pitch the most suitable and widely adopted parameter for representing numeric data in sonification.
Taking all these factors into consideration, we chose pitch as the baseline method in our study.

\subsection{Procedure}
Participants were asked to complete a total of 45 trials across four tasks (12 value comparison, 12 trend recognition, and 21 value sign and exact value identification trials).
The data value comparison and data trend recognition tasks each included three sets of questions with varying levels of difficulty, with four questions in each set, for a total of 12 questions. 
The value sign identification and exact value identification tasks shared the same set of 21 audio stimuli, covering all integers from $-10$ to $10$.
Each participant who completed the full experiment was compensated with 20 SGD as a token of appreciation for their time and involvement.

\textbf{Pre-Study Setup:} Before the test, participants were seated, and the study coordinator assisted them in adjusting the headphones to a comfortable volume. Participants were also given a brief explanation of the tasks and the two methods to be used.

\textbf{On-site Experiments:} \revision{To mitigate learning and fatigue effects, the order of conditions was counterbalanced.
Half of the participants performed tasks using SpaudioData first and then the baseline method, while the other half were tested on the two methods in the reverse order.} 
Each group underwent a training session before starting the tasks. 
After completing the tasks for the first method, \revision{participants were given a five-minute rest break before proceeding to the next method. The entire study took approximately 60 minutes for participants to finish all the tasks.} 

\subsection{Result}

\begin{figure*}
    \centerline{\includegraphics[width=\textwidth]{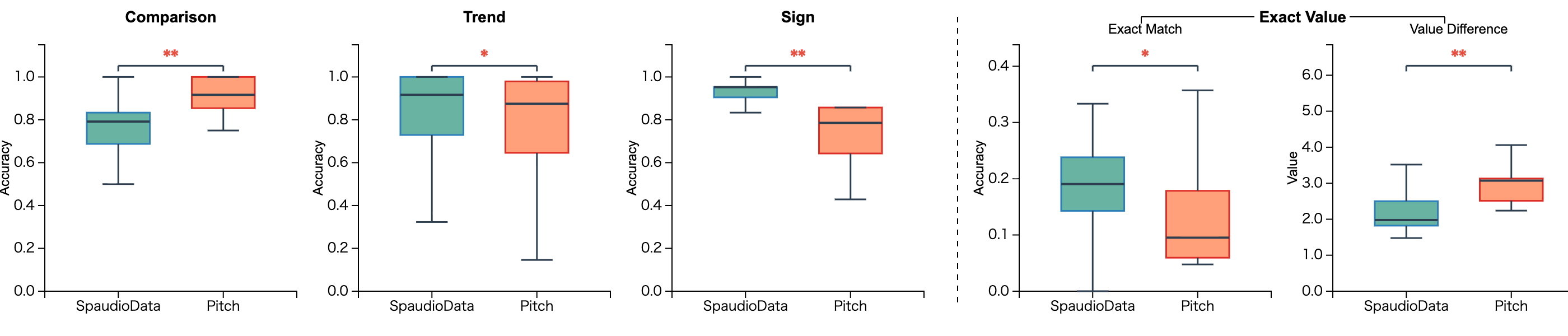}}
    \caption{Results of four tasks for BLV participants. Statistical significance is indicated by stars, where $^{***}$ denotes $p < 0.001$, $^{**}$ denotes $p < 0.01$, and $^{*}$ denotes $p < 0.05$, based on the Wilcoxon rank-sum test.
    }
    \label{fig:blind_participants}
\end{figure*}

\begin{figure*}
    \centerline{\includegraphics[width=\textwidth]{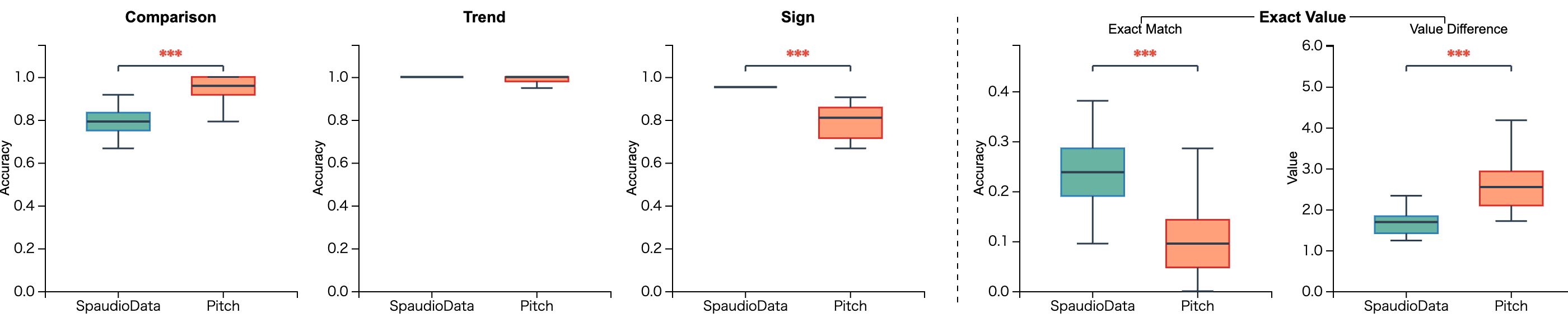}}
    \caption{Performance comparison between \toolName{} and the pitch representations for sighted participants across four tasks, showing consistent results with the overall participant group, with notably high accuracy and low variance in the data trend recognition tasks using \toolName{}.}
    \label{fig:sighted_participants}
\end{figure*}


In this study, we employed the Wilcoxon signed-rank test to analyze the performance differences between \toolName{} and pitch representation across four tasks.
We conducted analyses on overall performance, task-specific performance, and the influence of different data characteristics.
\revision{The Wilcoxon signed-rank test was chosen because the data did not meet the normality assumption required for parametric tests and the comparisons were paired within participants.}

\textbf{Analysis by Participant Group.}
\revision{As shown in \autoref{fig:blind_participants}, \toolName{} significantly outperformed the pitch representation in data trend recognition, sign identification and exact value identification.
\revisionminor{\toolName{} showed a statistically significant advantage in identifying exact value. The results in the exact match and difference value subplots confirm this, showing that \toolName{} produced more accurate responses and minimized the numerical discrepancy (i.e., lower difference values) compared to the pitch representation.}
However, when users were required to compare specific data values, the pitch representation showed higher performance, likely because pitch naturally supports relative magnitude comparison across data points.
For sighted participants as depicted in \autoref{fig:sighted_participants}, no significant difference was observed in the data trend recognition task, as both representations achieved accuracy levels close to 100\%, indicating a clear ceiling effect.
}

\textit{\textbf{Observation}: 
\revision{Compared to the pitch representation, for BLV participants, \toolName{} performs better in data trend recognition, exact value identification and sign identification tasks and shows lower effectiveness in data value comparison tasks.}
}


\begin{figure}[!htbp]
    \centerline{\includegraphics[width=\columnwidth]{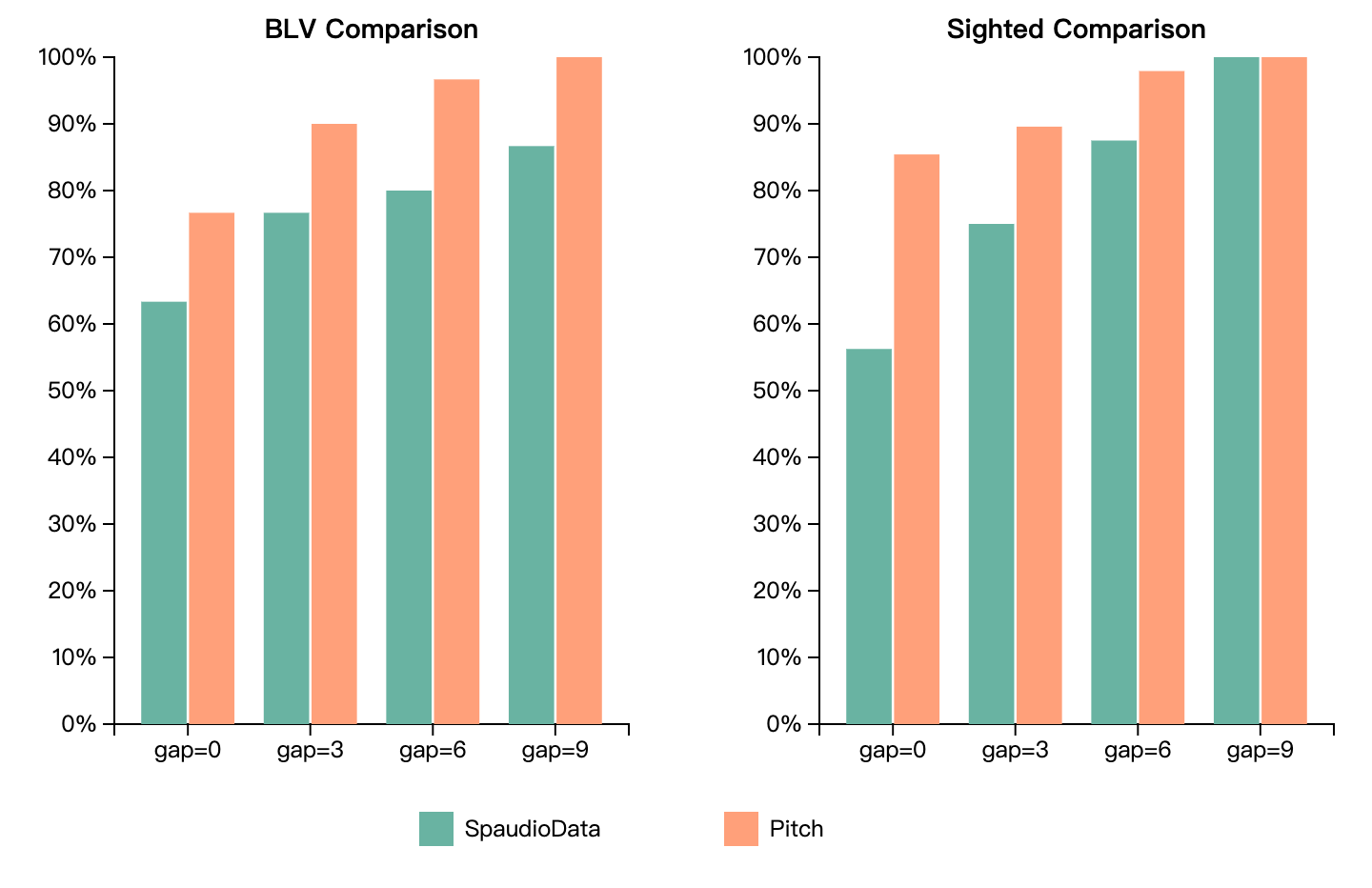}}
    \caption{Distribution of accuracy by different gaps for data value comparison task.
    \revisionminor{For the Comparison, Trend, Sign, and Exact Match tasks, higher values indicate better performance, whereas for the Value Difference task, lower values indicate better performance.}
    }
    \label{fig:gap_diff}
\end{figure}

\begin{figure}[!htbp]
    \centerline{\includegraphics[width=\columnwidth]{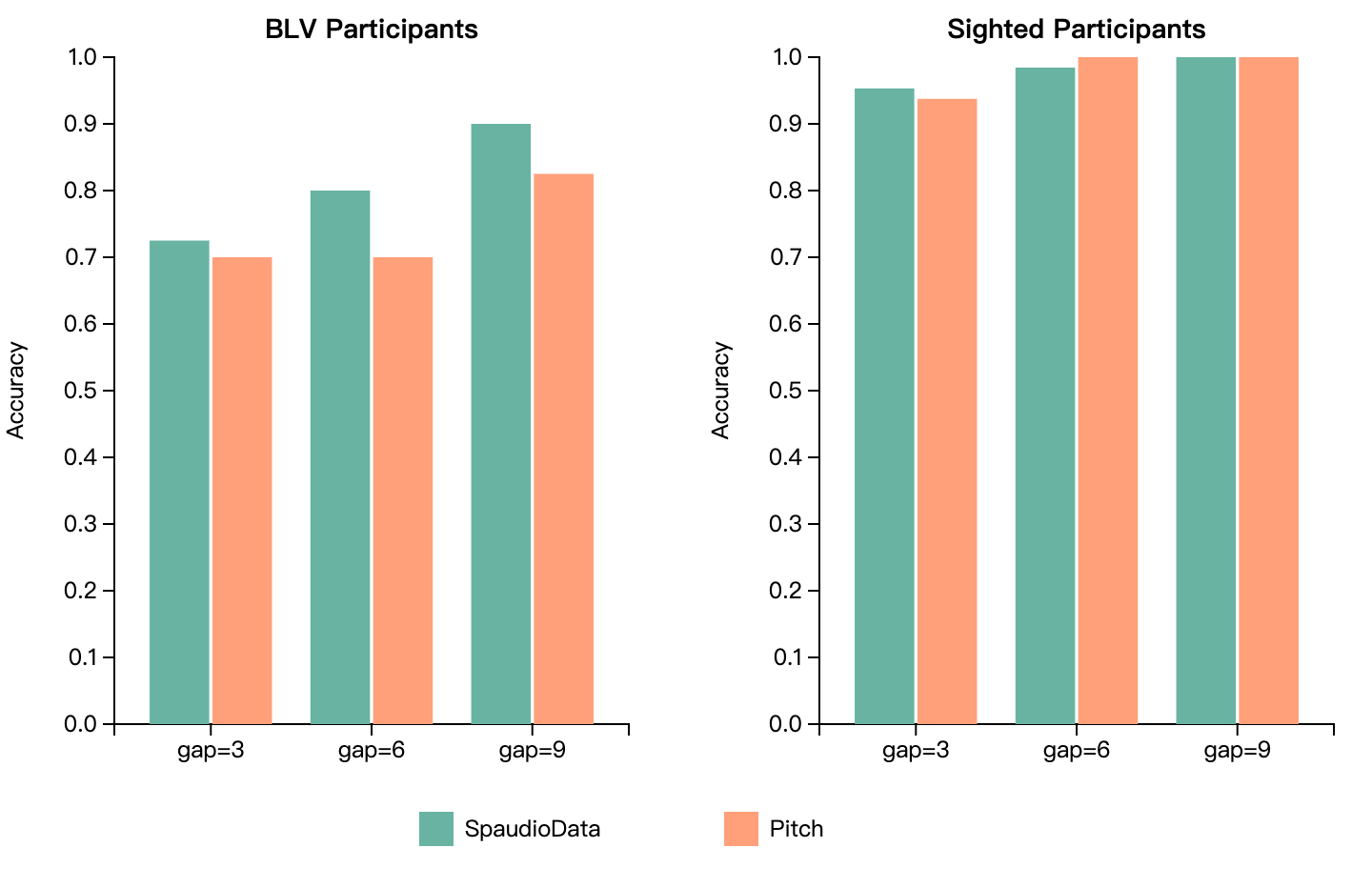}}
    \caption{Distribution of accuracy by different gaps for data trend recognition task.}
    \label{fig:gap_diff_trend}
\end{figure}

\textbf{Analysis by Task and Data Characteristics.}
In the data value comparison and data trend recognition tasks, the dataset was designed with varying gaps between numerical values.
We investigated how these gaps influenced participants’ data perception accuracy.
\revision{As shown in \autoref{fig:gap_diff}, in the data value comparison task, accuracy consistently increased as the gap increased from 0 to 9.
Larger gaps made the difference between the two values more salient, allowing participants to more reliably identify the larger one.
This pattern held for both BLV and sighted participants, indicating that greater numerical separation generally improves perceptual discriminability.
Across all gap conditions, \toolName{} consistently achieved higher accuracy than the pitch-based representation.
This result reflects the fact that BLV individuals often leverage auditory spatial cues, which might enhance their performance using \toolName{}.}
In the data trend recognition task shown in \autoref{fig:gap_diff_trend}, accuracy improved for both representations as the gap increased, indicating that larger numerical differences made trends easier to perceive.
Notably, the effect of representation differed across participant groups.
For BLV participants, \toolName{} led to consistently higher accuracy than the pitch-based representation, suggesting that its spatialized and multi-dimensional auditory cues better supported trend perception.
For sighted participants, the performance difference between \toolName{} and the pitch representation was relatively small.

\textit{\textbf{Observation}: Performance in both representations improves with larger numerical gaps, though \toolName{} shows a distinct advantage for BLV participants when comparing identical values.}

\begin{figure}[!htbp]
    \centerline{\includegraphics[width=\columnwidth]{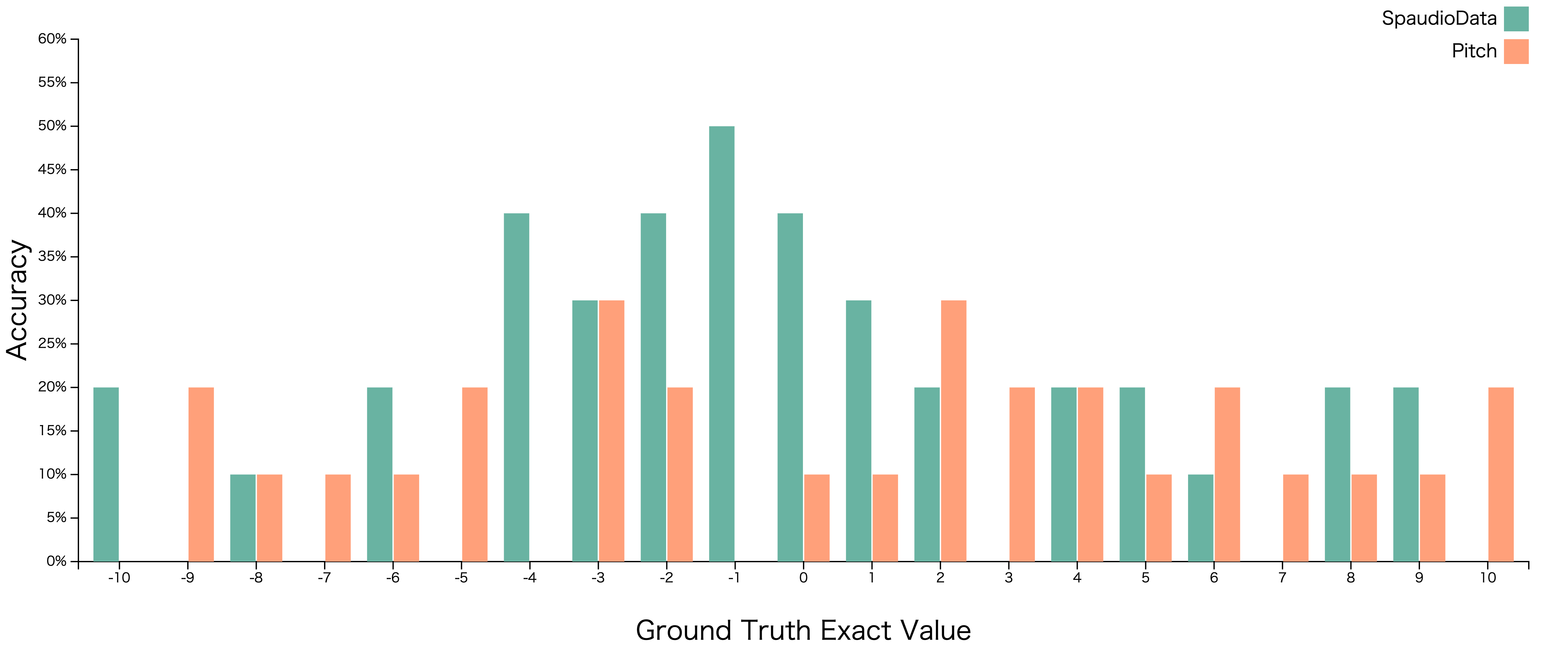}}
    \caption{\revision{Accuracy distribution across ground truth values for BLV participants.}}
    \label{fig:distribution_by_number_blv}
\end{figure}

\begin{figure}[!htbp]
    \centerline{\includegraphics[width=\columnwidth]{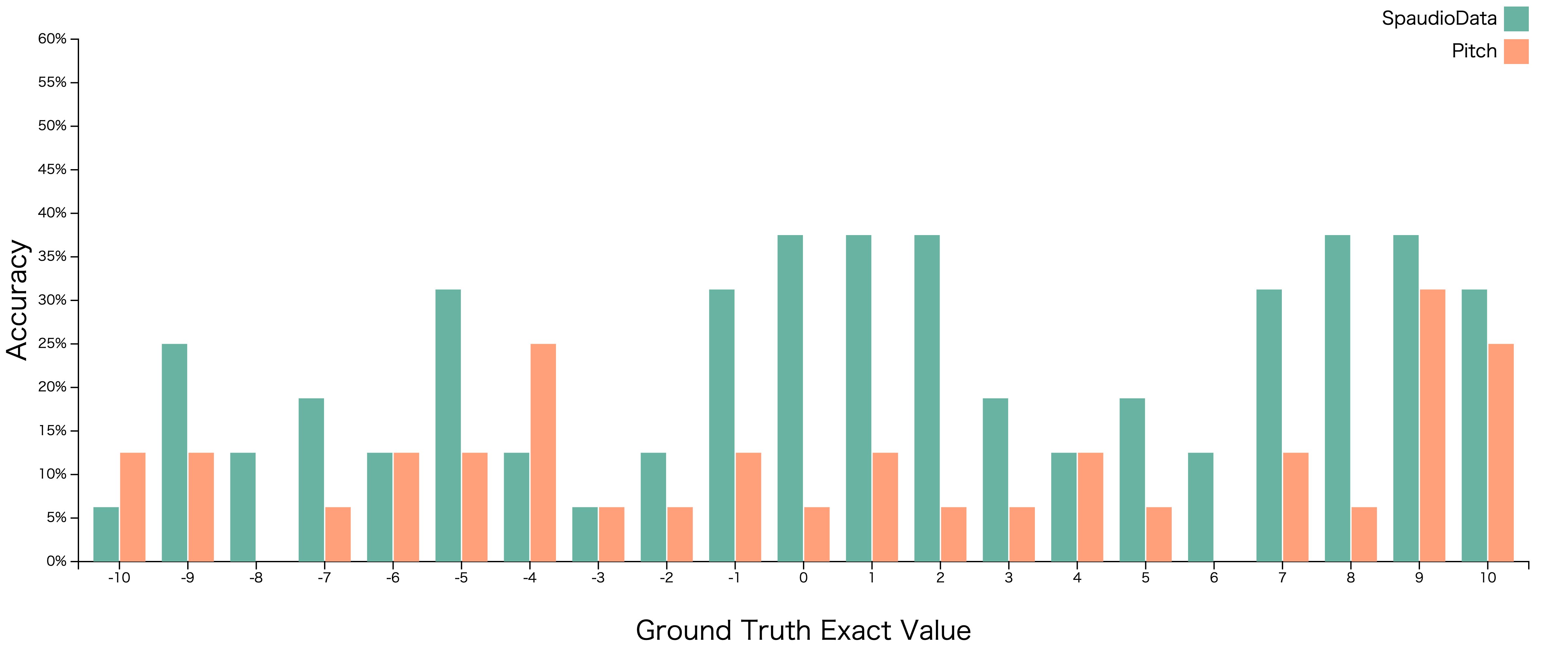}}
    \caption{\revision{Accuracy distribution across ground truth values for sighted participants.}}
    \label{fig:distribution_by_number_sighted}
\end{figure}
\textbf{Accuracy Distribution of Exact Match Across the Whole Range.}
For exact value identification task, \toolName{} consistently outperformed the pitch representation.
We further analyzed the accuracy of the two methods for different ground truth values (ranging from -10 to 10).
The results in \autoref{fig:distribution_by_number_blv} and \autoref{fig:distribution_by_number_sighted} show that \toolName{} achieved higher accuracy across most numerical ranges for both BLV and sighted participants.
Notably, accuracy was particularly high when the numbers were close to $0$, suggesting that participants were more sensitive to \toolName{} cues for sounds originating from central, frontal positions.
This heightened sensitivity may be attributable to everyday listening experiences, as sounds originating directly in front of the listener are typically more salient and easier to localize.

\textit{\textbf{Observation}: In the exact value identification task, \toolName{} demonstrates superior accuracy across most numerical ranges compared to the pitch representation, with particularly high performance for values near zero.}

\section{Discussion}
Our formal user study has demonstrated the effectiveness of \toolName{} for data representation, especially its significantly better performance than the pitch representation in the sign and exact value identification tasks, and comparable performance on the trend recognition task. However, \toolName{} is also inferior to the pitch representation in the value comparison task. We will further discuss the implications and possible limitations of our approach.

\textbf{Performance difference between data value comparison and exact value identification.} As reported above, \toolName{} performs significantly better than the pitch representation in the task of exact value identification, but is also significantly worse than the pitch representation in the data value comparison task, as shown in \autoref{fig:blind_participants} and \autoref{fig:sighted_participants}.
Given that exact value identification is generally believed to be more challenging than data value comparison, this result looks contradictory but actually makes sense. Prior perception research~\cite{hu2020comparative} has shown that people can easily recognize the difference of pitch. The lower accuracy of \toolName{} than the pitch representation shows that human users are less sensitive to sound direction than pitch, i.e., \textit{\textbf{the Just Noticeable Difference (JND)\footnote{\url{https://en.wikipedia.org/wiki/Just-noticeable_difference}} of sound direction is higher than that of pitch}}, which is also supported by the higher accuracy variance of \toolName{} regarding the change of value gaps (\autoref{fig:gap_diff}).
However, for the task of exact value identification, the performance depends on both the JND of the audio encoding channel and the accurate mapping from the audio encoding channel and the actual data value. Given that common users without professional musical training can easily recognize the sound direction, but it is often hard for them to identify the exact pitch value (e.g., in Hz) of a sound. It makes the mapping between sound direction and data value be more easily and accurately identified than pitch, which, we speculate, leads to the significantly better performance of \toolName{} than pitch in the task of exact value identification.

\textbf{Good balance between showing coarse-grained information and fine-grained data details.} \toolName{} can achieve significantly better performance than the pitch representation on representing fine-grained details like value sign and exact value, and still perform well on revealing coarse-grained data information. For example, \toolName{} achieved a mean accuracy of more than 90\% on the trend recognition task and about 80\% on the data value comparison task for BLV participants (\autoref{fig:blind_participants}).
We argue that this is actually a good balance for using sonification for data representation. With \toolName{}, users can perceive the overall data information and exact data value simultaneously without the need to read out exact values of individual numbers on demand via speech, which is currently a common practice in accessible data representation.

\textbf{Limitations.} While \toolName{} shows promising results for fine-grained data sonification, several limitations remain. First, our approach relies on spatial audio rendered through Head-Related Transfer Functions (HRTFs), which vary across individuals due to differences in head, ear, and torso morphology~\cite{yang2022audio,correa2023spatial}. Personalized HRTFs can improve spatial localization but require complex calibration, which is currently impractical for most users. 
\revision{
As noted in the Background section, using generalized HRTFs instead of personalized ones was a design choice to ensure accessibility.}
We employed Apple AirPods, which use generic HRTFs, to ensure acceptable spatial audio quality across participants.
Second, \toolName{} requires headphones or earphones capable of rendering spatial audio. Although this may limit accessibility in certain contexts, spatial audio hardware is becoming increasingly available and affordable, making this constraint less restrictive in practice.
Third, since we encode values using azimuthal direction (ranging from \(-90^\circ\) to \(90^\circ\)), there exists an inherent upper bound on the number of distinguishable data points. However, this resolution is sufficient for many everyday accessibility scenarios.
Finally, our user study, while covering four perceptual tasks (trend identification, comparison, sign recognition, exact value), does not encompass all real-world sonification use cases. Scenarios involving dynamic data streams, multivariate contexts, or exploratory interactions remain unexplored. Moreover, \revision{our study involved only 10 BLV participants, which is a relatively small size of participants.}
There were also noticeable demographic differences between the BLV and sighted participant groups: the BLV participants were generally older (\revision{average 49.7 years}),
whereas sighted participants were \revision{average} 25.5 years. The sighted group had, on average, a higher level of formal education. 
\revision{These factors in our study may have an influence on the auditory perception, task comprehension and response behaviors. A more in-depth and rigorous study on these factors is left as future work.}

\textbf{Informing Future Research.} 
Our study results confirm the JND differences between sound direction and pitch, which actually reveals the importance of investigating data representation effectiveness of different audio encoding channels. It is still an underexplored research direction, as pointed out by the prior survey on accessible data visualization~\cite{kim2021accessible}. Our work can inform future research on this direction. Also, given the different advantages of \toolName{} and the pitch representation, it is also interesting to explore the combination of pitch and sound direction, i.e., using them for the ``double-coding'' of the input data, to achieve a highly accurate representation of both coarse-grained and fine-grained data information via audio.





\section{Conclusion and Future Work}

In this paper, we propose a novel approach \toolName{} by using spatial audio to achieve accessible fine-grained data representation, which can benefit BLV individuals. Specifically, we use the sound direction in the azimuth plane to represent the data.
We conduct detailed user studies with \revision{26} participants (\revision{10} BLV participants) to evaluate the performance of \toolName{} by comparing it with the pitch representation across both coarse-grained and fine-grained tasks.
Our findings indicate that the widely used pitch representation method, while useful, often falls short in conveying the fine-grained detailed information of the data that is necessary for a comprehensive understanding and exploration of the input data. In contrast, \toolName{} can achieve significantly better performance on fine-grained data exploration tasks, i.e., value sign identification and exact value identification, and achieve an accuracy of more than 90\% in the data trend recognition task, and an inferior but still good (an accuracy of about 80\%) performance in the data comparison tasks. It demonstrates the \revision{effectiveness} of \toolName{} for accessible data representation.


In future work, we plan to combine sound direction and pitch to do ``double-coding'' of data via sound, and investigate its performance. Also, it is interesting to conduct in-depth research on the data representation effectiveness of different audio encoding channels, guiding the future research on accessible data representation via sonification.


\section{Acknowledgments}
The authors thank Zhengyang Ma and Yanna Lin for their valuable discussions during the course of this project.
This project is supported by the Ministry of Education, Singapore, under its Academic Research Fund Tier 2 (Proposal ID: T2EP202220049), and by NTU Start Up Grant awarded to Yong Wang. Any opinions, findings and conclusions, or recommendations expressed in this material are those of the author(s) and do not reflect the views of the Ministry of Education, Singapore.

\bibliographystyle{IEEEtran}
\bibliography{template}

\section{ABOUT THE AUTHORS}

\textbf{Can Liu} is currently a research fellow at the College of Computing and Data Science, Nanyang Technological University. 
He received his Ph.D. degree from Peking University in 2023. 
His research interests lie in AI-driven data visualization and human-computer interaction, especially intelligent data interaction, intelligent visual design, and intelligent data management.
For more details, please refer to http://vvoliucano.github.io/.

\quad \textbf{Wenjie Jiang} is currently a Ph.D. student at the College of Computing and Data Science, Nanyang Technological University. 
He received his B.Sc. degree in Computer Science from Southeast University. 
Wenjie’s research interests include data visualization and data summarization. 
Contact him at wenjie008@e.ntu.edu.sg.

\textbf{Shaolun Ruan} is currently a Ph.D. candidate in
School of Computing and Information Systems at
Singapore Management University (SMU). His work
focuses on developing novel graphical representations that enable a more effective and smoother
analysis for humans using machines, leveraging
the methods from Data Visualization and Humancomputer Interaction. He received his bachelor’s
degree from the University of Electronic Science and
Technology of China (UESTC) in 2019. For more
information, kindly visit https://shaolun-ruan.com/.

\textbf{Kotaro Hara} is an Assistant Professor of Computer Science in the School of Computing and Information Systems at Singapore Management University. He received his Ph.D. degree from the University of Maryland in 2016. His teaching topics include human-computer interaction and interaction design \& prototyping. His research areas and expertise cover human-machine collaborative systems, learning and work, urban mobility and smart environments, and health and wellbeing, with a focus on ageing and mental health. Contact him at kotarohara@smu.edu.sg.

\textbf{Yong Wang} is currently an assistant professor in the College of Computing and Data Science, Nanyang Technological University. Before that, he worked as an assistant professor at Singapore Management University from 2020 to 2024. His research interests include information visualization, visual analytics and human-AI collaboration, with an emphasis on their application to FinTech, quantum computing and online learning. He obtained his Ph.D. in Computer Science from Hong Kong University of Science and Technology. He received his B.E. and M.E. from Harbin Institute of Technology and Huazhong University of Science and Technology, respectively. For more details, please refer to http://yong-wang.org.

\end{document}